\documentclass{article}

\usepackage{amsmath}
\usepackage{amssymb}
\usepackage{graphicx}
\usepackage{epsfig}

\begin{document}
\title{A New Approach for Solving Singular Systems in Topology Optimization Using Krylov Subspace Methods}
\author{Teruyoshi WASHIZAWA$^{1}$，
Akira ASAI$^{1}$，and
Nobuhiro YOSHIKAWA$^{2}$\hfil \\[0.1\baselineskip] \normalsize
\hbox{\vbox{\flushleft
\hspace{3cm}$\quad^1$ Canon Inc. Research Center\\[-0.1\baselineskip] \normalsize
\hspace{3cm}$\quad^2$ Institute of Industrial Science, University of Tokyo\\ }}
}
\maketitle

\begin{abstract}
In topology optimization, the design parameter of the element that does not give any contribution to the objective function vanishes. This causes the stiffness matrix to become singular. To avoid the breakdown caused by this singularity, the previous studies employ some additional procedures. These additional procedures, however, have some problems. On the other hand, convergences of Krylov subspace methods as the solution of singular systems have been studied recently. By the consecutive studies, it has been revealed that conjugate gradient method (CGM) does not converge to the local optimal solutions in some singular systems but in those satisfying some condition while conjugate residual method (CRM) converges in any singular systems. In this article, we will show that a local optimal solution of topology optimization is obtained by using the CRM and the CGM as a solver of the equilibrium equation even if the stiffness matrix becomes singular. Moreover, we prove that the CGM without any additional procedures converges to a local optimal solution in that case. Computer simulation shows that the CGM gives approximately the same solutions obtained by the CRM in the case of a typical cantilever beam problem.
\end{abstract}

\section{Introduction}
In topology optimization problems, we try to obtain the shape and topology of the structure that minimizes a given objective function under given constraints.
The topology of the structure to be obtained in such problems may include the number of holes.
Although the homogenization method \cite{Bendsoe:Kikuchi1988} is one of the most effective methods for topology optimization in two-dimensional problems, its computational cost becomes much larger in three-dimensional problems.
For this reason, this article treats the material distribution method \cite{Bendsoe:Sigmund2003}, which has recently attracted attentions at the computational point of view.
In the material distribution method, the shape of the structure is represented as a density distribution.
The density distribution takes the value one for the element filled with material and zero for that containing no material.

\begin{figure}[hbtp]
  \begin{center}
  \includegraphics[scale=0.5]{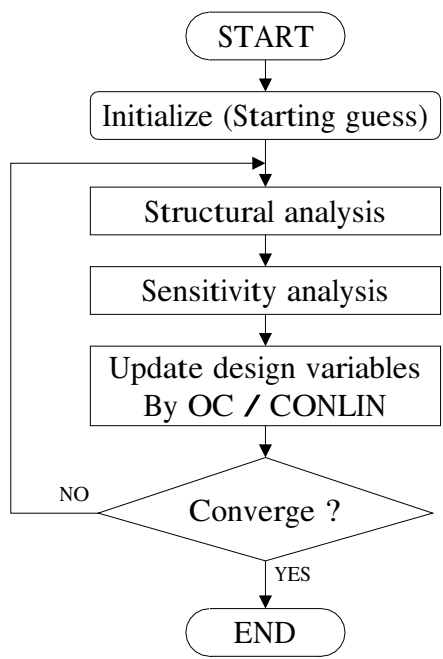}
    \caption{The flow of computations for topology design using the material distribution method and the OC / CONLIN methods for optimization.}
    \label{Flow01}
  \end{center}
\end{figure}

Topology optimization is solved by the following steps ( See Fig.\ref{Flow01} ):
\begin{enumerate}
\item initialize the density distribution of the structure.
\item repeat the following steps until convergence.
\item solve the structural analysis problem.
\item calculate the sensitivity of the density by the sensitivity analysis.
\item update the density distribution of the structure.
\end{enumerate}
In this kind of problems, it is always supposed that the design region is divided by finite uniform sized mesh, especially the finite element method (FEM) is employed in the structural analysis problem.
Then the density distribution, the displacement field, and the stress field defined on real space are described as finite dimensional vectors.

The densities of some elements of the structure often take the value zero in topology optimization.
This results in the singular stiffness matrix and the numerical breakdown of linear solvers in structural analysis.
The previous methods avoid this breakdown by using the following procedures:
\begin{enumerate}
\item reconstruct the system of equations to regularize the stiffness matrix.
\item restrict the range of the value of the density $\rho_j$ within $0 < \rho_{min} \le \rho_j \le 1$ to regularize the stiffness matrix.
$\rho_{min}$ is usually set to be $10^{-3}$ in typical applications \cite{Bendsoe:Sigmund2003}.
\end{enumerate}
These procedures, however, have some problems.
The first one causes the increment of the computational cost because the additional steps to find the elements with zero density and to reconstruct the system of equations are required.
The second one causes the  physical inconsistency so that the thin or weak material exists in the element that contains essentially no material.

In this article, we will show that such singular systems in topology optimization can be solved by Krylov subspace methods.
By these methods without any additional procedures, the value of $\rho$ is not restricted to be zero.
Especially we are concerned with conjugate residual method (CRM) and conjugate gradient method (CGM), the fundamental ones of the Krylov subspace methods.
We will show that a local optimal solution of topology optimization is obtained by using the CRM and the CGM as a solver of the equilibrium equation even if the stiffness matrix becomes singular.

The algorithm using the CGM as a solver of the structural analysis problem has already proposed by Fujii {\it et.al.} \cite{Fujii:Suzuki:Ohtsubo2000}
and described its merits for the computational cost.
However, they used the second additional procedure in accordance with the voxel FEM.
The convergence of the CGM for the singular stiffness matrix is not considered in their paper.
Besides, in CONLIN method \cite{Fleury1989}, well known as a solver of topology optimization problems, the CGM is used to obtain the updates of the Lagrange multipliers because singularity \cite{Fleury1989}.
As the reason for using the CGM as a solver, they only described that the CGM does not need the inverse of the stiffness matrix.
No discussion of convergence of the CGM to the local optimal solution is also given.

The convergence of the Krylov subspace methods for a system of linear equations with a singular coefficient matrix has been studied recently in numerical mathematics.
By this consecutive studies, it has been revealed that the CGM does not converge to the local optimal solutions in some singular systems but in those satisfying some condition \cite{Kaasschieter1988} and the conjugate residual method (CRM) converges to the local optimal solutions in any singular systems \cite{Abe:Ogata:Sugihara:Zhang:Mitsui1999, Hayami2001-1, Zhang:Oyanagi:Sugihara2000}.
Moreover, we will prove that the CGM without any additional procedures converges to a local optimal solution in that case.

The remaining part of this article is organized as follows.
First, topology optimization problem is formulated followed by the description of the convergence of the Krylov subspace methods.
In the section of the Krylov subspace methods, we give a brief introduction of the convergence theorems for the CRM and the CGM for regular systems and then consider their behaviors for singular systems based on the method introduced by Abe {\it et.al.} \cite{Abe:Ogata:Sugihara:Zhang:Mitsui1999}.
From this consideration, the sufficient condition of the convergence of the CGM for singular systems is obtained.
After above generic discussion, the convergence of the CGM in topology optimization is proved.
Computer simulation for the coat-hanging problem verifies our proof.

\section{Formulation of Topology Optimization}
Since topology optimization problems are described in detail in Refs. \cite{Bendsoe:Sigmund2003, Fujii2002, Fujii:Suzuki:Ohtsubo2000},
we give the brief introduction here.
Besides, we restrict the problem to a two dimensional plane strain problem, and suppose to formulate by using FEM for comprehensive description.

\begin{figure}[hbtp]
  \begin{center}
  \includegraphics[scale=0.5]{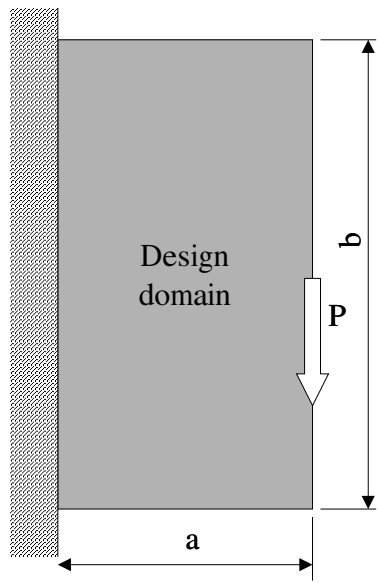}
    \caption{An example of the topology optimization : a clamped plate. A design domain is depicted as a shaded rectangle.}
    \label{Problem01}
  \end{center}
\end{figure}

The system we are concerned with is, for example, a cantilever with completely fixed left end and free right end with a concentrated load at the middle point shown in Fig.\ref{Problem01}.
The topology and size of the cantilever is determined as the solution of the minimization of a given objective function within a given domain.
This domain is called design domain and depicted as a shaded rectangle in Fig.\ref{Problem01}.
Now we formulate the topology optimization in accordance with the Refs. \cite{Bendsoe:Sigmund2003, Fujii:Suzuki:Ohtsubo2000}.

Based on the voxel FEM, a design domain is divided by a uniform sized mesh.
The material distribution (MD) method is employed for the representation of the shape of the structure.
In the MD method, the density $0 \le \rho \le 1$ is assigned to each element.
The density takes the value zero for the element with no material and one for the element with material.
A vector composed of all densities is called density vector $\rho$.
Assuming the density directly affects the stiffness of the element,
Young's modulus is a function of the density.
Therefore, every element of the total stiffness matrix $A$ is a function of the density.
The element stiffness matrix $A_j$ of the $j$-th element is also a function of the density $\rho_j$ as follows:
\begin{equation}
A_j = \rho_j^p D_j
\label{yousogouseigyouretu}
\end{equation}
where $p$ is a design parameter often set to be 2 or 3 and $D_j$ is an element stiffness matrix seen in ordinary structural analysis problems.
This formulation is known as Solid Isotropic Material with Penalization (SIMP) model \cite{Bendsoe:Sigmund2003}.

A displacement vector is assigned at every node of the mesh.
A rearranged vector composed of the displacement vector at every node is called a nodal displacement vector afresh and denoted as ${\bf x}$.

The density vector is determined so that the objective function is minimized under the given constraints.
The objective function is often defined as the mean compliance of the structure and the total volume ( or total mass ) and the lower and the upper bound of the density are used as the constraints as follows:
\begin{eqnarray}
\mbox{min} & & C(\rho) := \frac{1}{2} {\bf x}^T A(\rho) {\bf x}, \label{compliance} \\
\mbox{ subject to} & & \sum_j \rho_j \le \rho_0, \\
& & 0 \le \rho_j \le 1
\label{constraints}
\end{eqnarray}
The first derivative of the objective function with respect to the design variables is called sensitivity.
The sensitivity $C_{\rho_j}$ of Eq.(\ref{compliance}) is given as follows:
\begin{equation}
C_{\rho_j} := \frac{\partial C}{\partial \rho_j} = - {\bf x}^T \frac{\partial A}{\partial \rho_j} {\bf x}
\label{sensitivity0}
\end{equation}
Note that ${\bf x}$ is also the function of $\rho$ because $x$ is obtained as the solution of the linear equation $A(\rho) {\bf x} = {\bf b}$.
Here, ${\bf b}$ is a vector and the $j$th element $b(j)$ is an external force on the $j$th node, called nodal force.
Considering Eq.(\ref{yousogouseigyouretu}), we have\cite{Bendsoe:Sigmund2003, Fujii:Suzuki:Ohtsubo2000}
\begin{equation}
C_{\rho_j} = - {\bf x_j}^T (p \frac{A_j}{\rho_j} ) {\bf x_j}
\label{sensitivity}
\end{equation}
where $A_k$ is the element stiffness matrix of the $k$-th element.
${\bf x_k}$ is defined as a vector composed of the elements of the displacement vectors at the nodes belonging to the $k$-th element.
The dimensionality of ${\bf x_k}$ is, therefore, 8 in 2-D space and 24 in 3-D space.
The flow of the computation for the topology design is shown in Fig.\ref{Problem01}(b).
The density vector is obtained by the methods for optimization problem with constraints, i.e., optimality criteria (OC) method and convex linearization (CONLIN) method.
Both the OC and the CONLIN methods need the sensitivity in Eq.(\ref{sensitivity}).
The sensitivity is calculated in the "Sensitivity analysis" following the calculation of the displacement vector $x_j$ in the "Structural analysis" in Fig.\ref{Problem01}(b).

The density of the element that gives no contribution to the objective function will vanish.
It causes the total stiffness matrix to become singular.
In such a case, the previous methods avoid the numerical breakdown by using the following procedures:
\begin{enumerate}
\item reconstruct the system of equations to regularize the stiffness matrix.
\item restrict the range of value of the density to be $0 < \rho_{min} < \rho_j \le 1$.
$\rho_{min}$ is usually set to be $10^{-3}$ in typical applications \cite{Bendsoe:Sigmund2003}.
\end{enumerate}
These procedures, however, have some problems.
The first one causes the increment of the computational cost because the additional steps to find the elements with zero density and to reconstruct the system of equations are required.
The second one causes the  physical inconsistency so that the thin or weak material exists in the element that contains essentially no material.

\section{Krylov Subspace Methods in Regular Systems}
Before the discussion of the behavior of Krylov subspace method for singular systems,
we describe the behavior for regular systems.

\subsection{Conjugate Gradient Method}
When a matrix $A$ is positive definite, the solution ${\bf x}^*=A^{-1}{\bf b}$ of an equation $A{\bf x}={\bf b}$ is characterized as the minimum point of the following quadratic function:
\begin{eqnarray}
\phi(x) & = & \frac{1}{2}({\bf x}-{\bf x}^*,A({\bf x}-{\bf x}^*)) \\
& = & \frac{1}{2}({\bf x},A{\bf x})-({\bf x},b)+\frac{1}{2}({\bf x}^*,A{\bf x}^*)
\end{eqnarray}
Indeed, because of the positive definiteness of $A$,
$\phi({\bf x}) \ge 0 = \phi({\bf x}^*)$ is satisfied.
Therefore, by creating the sequence of the vectors ${\bf x}_0$，${\bf x}_1$，${\bf x}_2$，$\cdots$ which decreases $\phi({\bf x})$, it can be expected to obtain the approximate of the true solution ${\bf x}^*$.
This is the essential idea of the CGM.

The following is one of the CG algorithms:
\begin{quote}
{\em CG Algorithm}\cite{Mori:Sugihara:Murota1994}
\begin{enumerate}
\item Set the initial guess ${\bf x}_0$. Similarly, 
\begin{eqnarray}
{\bf r}_0 & := & {\bf b} - A{\bf x}_0, \\
{\bf p}_0 & := & {\bf r}_0
\end{eqnarray}
\item For $k=0,1,\cdots$, repeat the following steps until $\| {\bf r}_k \| \le \epsilon \| {\bf b} \|$ where $\epsilon > 0$ is a predetermined value:
\begin{eqnarray}
\alpha_k & = & \frac{({\bf r}_k,{\bf p}_k)}{({\bf p}_k,A{\bf p}_k)}, \label{CG01} \\
{\bf x}_{k+1} & = & {\bf x}_k + \alpha_k{\bf p}_k, \label{CG02} \\
{\bf r}_{k+1} & = & {\bf r}_k - \alpha_kA{\bf p}_k, \label{CG03} \\
\beta_k & = & - \frac{({\bf r}_{k+1},A{\bf p}_k)}{({\bf p}_k,A{\bf p}_k)}, \label{CG04} \\
{\bf p}_{k+1} & = & {\bf r}_{k+1} + \beta_k{\bf p}_k. \label{CG05}
\end{eqnarray}
\end{enumerate}
\end{quote}

{\em Theorem 3-1}\cite{Mori:Sugihara:Murota1994}
\begin{quote}
In the CG method for a symmetry positive definite matrix, the norm of the error $\| {\bf x}_k - {\bf x}^* \|$ decreases monotonically.
\end{quote}

\subsection{Conjugate Residual Method}
In the CRM, a half of the square of the norm of the residual is used as the objective function:
\begin{equation}
\tilde{\phi}({\bf x}) = \frac{1}{2}(A{\bf x}-{\bf b},A{\bf x}-{\bf b}) = \frac{1}{2}({\bf r},{\bf r})
\end{equation}
and obtain the approximate solution of $A{\bf x}={\bf b}$ by successive minimization.

The following is one of the CR algorithms:
\begin{quote}
{\em CR Algorithm}\cite{Mori:Sugihara:Murota1994}
\begin{enumerate}
\item Set the initial guess ${\bf x}_0$. Similarly,
\begin{eqnarray}
{\bf r}_0 & := & {\bf b} - A{\bf x}_0, \\
{\bf p}_0 & := & {\bf r}_0
\end{eqnarray}
\item For $k=0,1,\cdots$, repeat the following steps until $\| {\bf r}_k \| \le \epsilon \| {\bf b} \|$:
\begin{eqnarray}
\alpha_k & = & \frac{({\bf r}_k,A{\bf p}_k)}{(A{\bf p}_k,A{\bf p}_k)}, \label{CR01} \\
{\bf x}_{k+1} & = & {\bf x}_k + \alpha_k{\bf p}_k, \label{CR02} \\
{\bf r}_{k+1} & = & {\bf r}_k - \alpha_kA{\bf p}_k, \label{CR03} \\
\beta_k & = & - \frac{(A{\bf r}_{k+1},A{\bf p}_k)}{(A{\bf p}_k,A{\bf p}_k)}, \label{CR04} \\
{\bf p}_{k+1} & = & {\bf r}_{k+1} + \beta_k{\bf p}_k. \label{CR05}
\end{eqnarray}
\end{enumerate}
\end{quote}
The following theorem was proved for evaluating the convergence of the CRM \cite{Hayami2001-1}：\\
{\em Theorem 3-2}
\begin{quote}
If the symmetric part $M = (A+A^T)/2$ of the coefficient matrix $A$ is definite (positive or negative definite ), either of the following holds.
\begin{enumerate}
\item There exists $k^{\sharp} \ge 0$ so that 
${\bf p}_k \neq {\bf 0},(0 \le k \le k^{\sharp}-1)$ and ${\bf r}_{k^{\sharp}}=0$.
Further, the following relation holds for $0 \le k \le k^{\sharp}-1$:
\begin{equation}
\frac{\| {\bf r}_{k+1} \|^2}{\| {\bf r}_k \|^2} \le 1 - \frac{\{ \lambda_{min}(M) \}^2}{\lambda_{max}(A^TA)}
\label{CRConverge01}
\end{equation}
where $\lambda_{max}$，$\lambda_{min}$ is the maximum and minimum eigenvalue of the diagonal matrix, respectively.
\item ${\bf p}_k \neq 0$ and ${\bf r}_k \neq 0$ for all $k \ge 0$ and Eq.(\ref{CRConverge01}) holds.
\end{enumerate}
\end{quote}

The necessary and sufficient condition for the convergence without "breakdown" of the CRM for an arbitrary $b$ can be derived from the theorem 3-2.
The derived theorem also provides the meaning of the definiteness of $M$.
Note that "breakdown" means "the denominator of the parameter $\alpha_k$ in the conjugate residual algorithm becomes zero so that it becomes impossible to continue the computation."
The derived theorem said that the CRM converges without breakdown for an arbitrary ${\bf b}$ (for an arbitrary initial guess ${\bf x}_0$)\cite{Abe:Ogata:Sugihara:Zhang:Mitsui1999}.


\section{Krylov Subspace Methods in Singular Systems}
Now we examine the convergence of the Krylov subspace method for singular systems.
When dealing with singular systems, the range space $R(A)$ created by the row vectors of the coefficient matrix $A$ and the kernel ( or null space ) $N(A)$ spanned by ${\bf x}$ satisfying $A{\bf x}=0$ play the essential roles.

According to the analysis by Abe {\it et. al.} \cite{Abe:Ogata:Sugihara:Zhang:Mitsui1999}, we prepare the following variables:
\begin{eqnarray*}
& & r := rank(A) = dim(R(A)) > 0, \\
& & {\bf q}_1, \cdots, {\bf q}_r : \mbox{orthonormal basis of $R(A)$} \\
& & {\bf q}_{r+1}, \cdots, {\bf q}_n : \mbox{orthonormal basis of $R(A)^{\bot}$} \\
& & Q_1 := ( {\bf q}_1, \cdots, {\bf q}_r ) : \mbox{$n \times r$ matrix} \\
& & Q_2 := ( {\bf q}_{r+1}, \cdots, {\bf q}_n ) : \mbox{$n \times (n-r)$ matrix} \\
& & Q := (Q_1, Q_2) : \mbox{$n \times n$ orthogonal matrix}
\end{eqnarray*}
where $R(A)^{\bot}$ is the orthogonal complement of $R(A)$.
The following orthogonal transformation provides the representation of the Krylov subspace in above coordinates:
\begin{equation}
\tilde{A} = Q^T A Q = \left(
\begin{array}{cc}
A_{11} & A_{12} \\
0      & 0
\end{array}
\right)
\end{equation}
where $A_{11} := Q_1^TAQ_1$ and $A_{12} := Q_1^TAQ_2$.
By this orthogonal transformation, the matrix $A$ can be represented as $\tilde{A}$ in "standard system" where the structure of the system is more clear.

Then, denoting direct sum as $\oplus$, the followings hold:\\
{\em Theorem 4-1}\\
\hspace{1cm}
$A_{11} = Q_1^T A Q_1$ is regular $\Leftrightarrow$ $R(A) \oplus N(A) = {\mathbb R^n}$ \\
{\em Theorem 4-2}\\
\hspace{1cm}
$A_{12} = Q_1^T A Q_2 = 0$ $\Leftrightarrow$ $R(A)^{\bot} = N(A)$ \\
{\em lemma 4-1}\\
\hspace{1cm}
$R(A)^{\bot} = N(A)$ $\Leftrightarrow A_{11}$ is regular \\

From theorem 4-2, when the condition "$R(A)^{\bot} = N(A)$" is satisfied, the standard system reduces more simple structure as follows:
\begin{equation}
\tilde{A} = Q^TAQ = \left(
\begin{array}{cc}
A_{11} & 0 \\
0      & 0
\end{array}
\right)
\label{standard02}
\end{equation}
and, from lemma 4-1, $A_{11} = Q_1^TAQ_1$ is regular in this case.
In topology optimization, the partial stiffness matrix is regular even if the total stiffness matrix is singular, we can use Eq.(\ref{standard02}) as the standard system.

\subsection{Convergence Property of CRM}
It is guaranteed for the CRM to converge to the local optimal solution for a system of linear equations with a singular coefficient matrix \cite{Abe:Ogata:Sugihara:Zhang:Mitsui1999}．
For comparison with the CGM described later, we show the method of analysis described in \cite{Abe:Ogata:Sugihara:Zhang:Mitsui1999}.

The coefficients $\alpha_k$ and $\beta_k$ in the CRM is calculated as follows:
\begin{eqnarray*}
\alpha_k & = & \frac{({\bf r}_k,A{\bf p}_k)}{(A{\bf p}_k,A{\bf p}_k)}, \\
\beta_k & = & \frac{(A{\bf r}_{k+1},A{\bf p}_k)}{(A{\bf p}_k,A{\bf p}_k)}
\end{eqnarray*}
Using the standard system Eq.(\ref{standard02}), these coefficients are decomposed into the $R(A)$ component (with superfix $\|$) and $N(A)$ component (with superfix $\bot$),
\begin{eqnarray*}
\alpha_k
 & = & \frac{ \left(
\left( \begin{array}{c} {\bf r}_k^{\|} \\ {\bf r}_k^{\bot} \end{array} \right),
\left( \begin{array}{cc} A_{11} & 0 \\ 0 & 0 \end{array} \right)
\left( \begin{array}{c} {\bf p}_k^{\|} \\ {\bf p}_k^{\bot} \end{array} \right)
\right) }
{ \left(
\left( \begin{array}{cc} A_{11} & 0 \\ 0 & 0 \end{array} \right)
\left( \begin{array}{c} {\bf p}_k^{\|} \\ {\bf p}_k^{\bot} \end{array} \right),
\left( \begin{array}{cc} A_{11} & 0 \\ 0 & 0 \end{array} \right)
\left( \begin{array}{c} {\bf p}_k^{\|} \\ {\bf p}_k^{\bot} \end{array} \right)
\right) } \nonumber \\
& = & \frac{ ( {\bf r}_k^{\|}, A_{11} {\bf p}_k^{\|} ) }{ ( A_{11} {\bf p}_k^{\|}, A_{11} {\bf p}_k^{\|} ) }, \\
\beta_k
 & = & \frac{ \left(
\left( \begin{array}{cc} A_{11} & 0 \\ 0 & 0 \end{array} \right)
\left( \begin{array}{c} {\bf r}_{k+1}^{\|} \\ {\bf r}_{k+1}^{\bot} \end{array} \right),
\left( \begin{array}{cc} A_{11} & 0 \\ 0 & 0 \end{array} \right)
\left( \begin{array}{c} {\bf p}_k^{\|} \\ {\bf p}_k^{\bot} \end{array} \right)
\right) }
{ \left(
\left( \begin{array}{cc} A_{11} & 0 \\ 0 & 0 \end{array} \right)
\left( \begin{array}{c} {\bf p}_k^{\|} \\ {\bf p}_k^{\bot} \end{array} \right),
\left( \begin{array}{cc} A_{11} & 0 \\ 0 & 0 \end{array} \right)
\left( \begin{array}{c} {\bf p}_k^{\|} \\ {\bf p}_k^{\bot} \end{array} \right)
\right) } \nonumber \\
& = & \frac{ ( A_{11} {\bf r}_{k+1}^{\|}, A_{11}{\bf p}_k^{\|} ) }{ ( A_{11} {\bf p}_k^{\|}, A_{11} {\bf p}_k^{\|} ) }
\end{eqnarray*}
Using above equations, the algorithm of the CR method can be decomposed into the $R(A)$ and the $N(A)$ component.\\ \newline \newline
\rule{10cm}{0.2mm} \newline
{\bf The decomposed CR algorithm}
\begin{enumerate}
\item Choose ${\bf x}_0^{\|}$ and ${\bf x}_0^{\bot}$.
Similarly, 
\begin{eqnarray*}
\mbox{\underline{$R(A)$ component}} & & \mbox{\underline{$N(A)$ component}} \nonumber \\
{\bf r}_0^{\|} = {\bf b}^{\|} - A_{11} {\bf x}_0^{\|}, & & {\bf r}_0^{\bot} = {\bf b}^{\bot}, \\
{\bf p}_0^{\|} = {\bf r}_0^{\|}, & & {\bf p}_0^{\bot} = {\bf r}_0^{\bot}
\end{eqnarray*}
For \item $k=0,1,\cdots$, repeat the following steps until the $R(A)$ component of the residual converges:\\
\begin{eqnarray*}
\mbox{\underline{$R(A)$ component}} & & \mbox{\underline{$N(A)$ component}} \nonumber \\
\alpha_k = \frac{({\bf r}_k^{\|},A_{11}{\bf p}_k^{\|})}{(A_{11}{\bf p}_k^{\|},A_{11}{\bf p}_k^{\|})}, & & \\
{\bf x}_{k+1}^{\|} = {\bf x}_k^{\|} + \alpha_k {\bf p}_k^{\|}, & &
{\bf x}_{k+1}^{\bot} = {\bf x}_k^{\bot} + \alpha_k {\bf p}_k^{\bot}, \\
{\bf r}_{k+1}^{\|} = {\bf r}_k^{\|} + \alpha_k A_{11} {\bf p}_k^{\|}, & &
{\bf r}_{k+1}^{\bot} = {\bf r}_k^{\bot}, \\
\beta_k = \frac{({\bf r}_{k+1}^{\|}, A_{11}{\bf p}_k^{\|})}{({\bf p}_k^{\|},A_{11}{\bf p}_k^{\|})}, & & \\
{\bf p}_{k+1}^{\|} = {\bf r}_{k+1}^{\|} + \beta_k {\bf p}_k^{\|}, & &
{\bf p}_{k+1}^{\bot} = {\bf r}_{k+1}^{\bot} + \beta_k {\bf p}_k^{\bot}
\end{eqnarray*}
\end{enumerate}
\rule{10cm}{0.2mm} \newline \newline
Since the algorithm for the $R(A)$ component can be regarded as the CRM applied to the $R(A)$ subsystem, it is guaranteed that the norm of the $R(A)$ component of the residual decreases monotonically from theorem 3-3.
Since the $N(A)$ component of the residual is equal to ${\bf b}^{\bot}$ and unchanged, consequently for an arbitrary $b$ the convergence of the CRM for singular systems is guaranteed.
On the other hand, there has no information about ${\bf x}^{\bot}$.
However, when ${\bf b}^{\bot}=0$, ${\bf r}^{\bot}=0$ and ${\bf p}^{\bot}=0$ and then ${\bf x}^{\bot}$ is constant.

Because of the convergence of the CRM in singular systems as described above, we can use the CRM to obtain a local optimal solution in topology optimization even if the stiffness matrix becomes singular.
In terms of computational cost, however, the CGM will be much effective than the CRM.
Then we will examine the convergence of the CGM in singular systems.

\subsection{Convergence Property of CGM}
Similar to the previous section, we look the CGM in the standard system below.
The fundamental steps of the CGM are described in the previous section.

For analyzing the behavior of the $R(A)$ component and the $N(A)$ component of the CGM, first we decompose the vectors ${\bf x}$, ${\bf p}$, ${\bf b}$, and ${\bf r}$ into these subspaces.
\begin{eqnarray*}
\tilde{{\bf x}} & = & Q^T {\bf x}
			= ({\bf x}^{\|}, {\bf x}^{\bot})^T, \\
\tilde{{\bf p}} & = & Q^T {\bf p}
			= ({\bf p}^{\|}, {\bf p}^{\bot})^T, \\
\tilde{{\bf b}} & = & Q^T {\bf b}
			= ({\bf b}^{\|}, {\bf b}^{\bot})^T, \\
\tilde{{\bf r}} & = & Q^T {\bf r}
			= ({\bf r}^{\|}, {\bf r}^{\bot})^T
\end{eqnarray*}
Eqs.(\ref{CG02},\ref{CG03},\ref{CG05}) can be decomposed into the $R(A)$ and the $N(A)$ component as follows:
\begin{eqnarray*}
\begin{array}{ll}
{\bf x}_{k+1}^{\|} = {\bf x}_k^{\|} + \alpha_k {\bf p}_k^{\|}, & 
{\bf x}_{k+1}^{\bot} = {\bf x}_k^{\bot} + \alpha_k {\bf p}_k^{\bot}, \\
{\bf r}_{k+1}^{\|} = {\bf r}_k^{\|} - \alpha_k A {\bf p}_k^{\|}, & 
{\bf r}_{k+1}^{\bot} = {\bf r}_k^{\bot}, \\
{\bf p}_{k+1}^{\|} = {\bf p}_k^{\|} + \beta_k {\bf p}_k^{\|}, & 
{\bf p}_{k+1}^{\bot} = {\bf p}_k^{\bot} + \beta_k {\bf p}_k^{\bot}.
\end{array}
\end{eqnarray*}
Then $\alpha_k$ and $\beta_k$ can be rewritten as follows:
\begin{eqnarray}
\alpha_k & = & \frac{ \left(
\left( \begin{array}{c} {\bf r}_k^{\|} \\ {\bf r}_k^{\bot} \end{array} \right),
\left( \begin{array}{c} {\bf p}_k^{\|} \\ {\bf p}_k^{\bot} \end{array} \right)
\right) }
{ \left(
\left( \begin{array}{c} {\bf p}_k^{\|} \\ {\bf p}_k^{\bot} \end{array} \right),
\left( \begin{array}{cc} A_{11} & 0 \\ 0 & 0 \end{array} \right)
\left( \begin{array}{c} {\bf p}_k^{\|} \\ {\bf p}_k^{\bot} \end{array} \right)
\right) } \nonumber \\
& = & \frac{ ( {\bf r}_k^{\|}, {\bf p}_k^{\|} ) }{ ( {\bf p}_k^{\|}, A_{11} {\bf p}_k^{\|} ) } + \frac{ ( {\bf r}_k^{\bot}, {\bf p}_k^{\bot} ) }{ ( {\bf p}_k^{\|}, A_{11} {\bf p}_k^{\|} ) } \nonumber \\
& = & \alpha_k^{\|} + \alpha_k^{\bot}, \label{CGalpha} \\
\beta_k & = & \frac{ \left(
\left( \begin{array}{c} {\bf r}_{k+1}^{\|} \\ {\bf r}_{k+1}^{\bot} \end{array} \right),
\left( \begin{array}{cc} A_{11} & 0 \\ 0 & 0 \end{array} \right)
\left( \begin{array}{c} {\bf p}_k^{\|} \\ {\bf p}_k^{\bot} \end{array} \right)
\right) }
{ \left(
\left( \begin{array}{c} {\bf p}_k^{\|} \\ {\bf p}_k^{\bot} \end{array} \right),
\left( \begin{array}{cc} A_{11} & 0 \\ 0 & 0 \end{array} \right)
\left( \begin{array}{c} {\bf p}_k^{\|} \\ {\bf p}_k^{\bot} \end{array} \right)
\right) } \nonumber \\
& = & \frac{ ( {\bf r}_{k+1}^{\|}, A_{11}{\bf p}_k^{\|} ) }{ ( {\bf p}_k^{\|}, A_{11} {\bf p}_k^{\|} ) }
\label{CGbeta}
\end{eqnarray}
Using above descriptions, we try to decompose the CG method into the $R(A)$ and the $N(A)$ component.\newline \newline
\rule{10cm}{0.2mm} \newline
{\bf The behavior of the $R(A)$ and the $N(A)$ component of the CGM}
\begin{enumerate}
\item Choose ${\bf x}_0^{\|}$ and ${\bf x}_0^{\bot}$.
Similarly,
\begin{eqnarray*}
\mbox{\underline{$R(A)$ component}} & & \mbox{\underline{$N(A)$ component}} \nonumber \\
{\bf r}_0^{\|} = {\bf b}^{\|} - A_{11} {\bf x}_0^{\|}, & & {\bf r}_0^{\bot} = {\bf b}^{\bot}, \\
{\bf p}_0^{\|} = {\bf r}_-^{\|}, & & {\bf p}_0^{\bot} = {\bf r}_0^{\bot}
\end{eqnarray*}
\item For $k=0,1,\cdots$, repeat the following steps until the $R(A)$ component of the residual converges:
\begin{eqnarray*}
\mbox{\underline{$R(A)$ component}} & & \mbox{\underline{$N(A)$ component}} \nonumber \\
\alpha_k^{\|} = \frac{({\bf r}_k^{\|},{\bf p}_k^{\|})}{({\bf p}_k^{\|},A_{11}{\bf p}_k^{\|})}, & &
\alpha_k^{\bot} = \frac{({\bf r}_k^{\bot},{\bf p}_k^{\bot})}{({\bf p}_k^{\|},A_{11}{\bf p}_k^{\|})}, \\
{\bf x}_{k+1}^{\|} = {\bf x}_k^{\|} + \alpha_k {\bf p}_k^{\|}, & &
{\bf x}_{k+1}^{\bot} = {\bf x}_k^{\bot} + \alpha_k {\bf p}_k^{\bot}, \\
{\bf r}_{k+1}^{\|} = {\bf r}_k^{\|} + \alpha_k A_{11} {\bf p}_k^{\|}, & &
{\bf r}_{k+1}^{\bot} = {\bf r}_k^{\bot}, \\
\beta_k = \frac{({\bf r}_{k+1}^{\|}, A_{11}{\bf p}_k^{\|})}{({\bf p}_k^{\|},A_{11}{\bf p}_k^{\|})}, & & \\
{\bf p}_{k+1}^{\|} = {\bf r}_{k+1}^{\|} + \beta_k {\bf p}_k^{\|}, & &
{\bf p}_{k+1}^{\bot} = {\bf r}_{k+1}^{\bot} + \beta_k {\bf p}_k^{\bot}
\end{eqnarray*}
\end{enumerate}
\rule{10cm}{0.2mm} \newline

For the CGM for symmetric definite matrix, it is guaranteed that the norm of the error decreases monotonically from theorem 3-1.
Therefore, if the $R(A)$ component of the CG algorithm is closed in the $R(A)$ subspace, it is guaranteed that the norm of the error decreases monotonically.
Since the $N(A)$ component of the residual is equal to ${\bf b}^{\bot}$ and unchanged,
the convergence of the CGM for singular systems is guaranteed in such a case.
However, focusing on the coefficient $\alpha_k^{\bot}$, as described in Eq.(\ref{CGalpha}), we can see its numerator includes ${\bf r}_k^{\bot}$ and ${\bf p}_k^{\bot}$, the $N(A)$ components of ${\bf r}_k$ and ${\bf p}_k$.
Since the $R(A)$ component of the CGM is not closed in the $R(A)$ subspace, the convergence of the CGM for singular systems is not guaranteed.
Indeed, the numerical experiments that the CGM diverges when ${\bf b}^{\bot} \neq 0$ has been reported \cite{Kaasschieter1988}．
If ${\bf b} \in R(A)$, that is ${\bf b}^{\bot}=0$, then ${\bf r}_0^{\bot}=0$, ${\bf r}_k^{\bot}=0, ( k = 1, 2, \cdots )$, and $\alpha_k^{\bot} = 0, ( k = 1, 2, \cdots )$, resulting in the convergence of the $R(A)$ subspace.
This is the sufficient condition of convergence of the CGM for singular systems.

\section{Singular Systems in Topology Optimization}
When the densities of all elements adjacent to a node $j$ takes the value 0,
all the corresponding components of the total stiffness matrix vanish.
Then the total stiffness matrix becomes singular and numerical algorithms for solving the equilibrium problem might be break down.
In order to make the matrix regular, reconstructing the system of linear equations by removing the $j$-th row and the $j$-th column.
Since the reconstructed stiffness matrix is considered to be $A_{11}$ in Eq.(\ref{standard02}), the range space of the reconstructed stiffness matrix corresponds to $R(A)$ and the removed part $N(A)$.
If the $j$-th component of the nodal force vector ${\bf b}$ takes the value 0, then ${\bf b} \in R(A)$,
and from the discussion in the previous section it is guaranteed that the CGM converges without breakdown.

Now the proposition to be proved as the sufficient condition for the convergence of the CGM is as follows:\\
{\em Proposition-1}
\begin{quote}
{\em If the densities of all elements adjacent to the node $j$ takes the value 0, the nodal force at the node $j$ is 0.}
\end{quote}
To prove the proposition-1 directly is difficult because the proposition-1 must evaluate the nodal forces after giving the densities of the elements while the practical computational process calculates the densities after giving the nodal forces.
Then we prove the contraposition of the proposition-1:\\
{\em proposition-2}
\begin{quote}
{\em If the nodal force at the $j$-th node is not 0, there exists at least one element which density takes the positive value among those adjacent to the element $j$.}
\end{quote}

\subsection{Proof of Proposition-2}
Assuming $b(j) \neq 0$ for $A{\bf x}={\bf b}$, then
\begin{equation}
0 \neq b(j) = A(j,:) x = \sum_{k=1}^{k=n_j} A_{k}(j_{k},:)x_{k}
\end{equation}
where $k$ is an index of an element adjacent to a node $j$,
${\bf x}_k$ and $A_{k}$ are the element displacement vector and the element stiffness matrix of the $k$-th element, respectively.
$j_{k}$ is the local index of the $j$-th node among the nodes located on the boundary of the $k$-th element.
$A_{k}(j_{k},:)$ is the row vector corresponding to the $j_{k}$-th node of $A_{k}$.
Among $n_j$ terms in the left hand side of above equation, there exists at least one element, say the $\overline{k}$-th element, so that $x_{\overline{k}} \neq 0$ and $A_{\overline{k}}(j_{\overline{k}},:){\bf x}_{\overline{k}} \neq 0$.

The sensitivity $C_{\rho_{\overline{k}}}$ of $C(\rho)$ with respect to the density of the ${\overline{k}}$-th element is given from the eq.(\ref{sensitivity}):
\begin{equation}
C_{\rho_{\overline{k}}} = - {\bf x}_{\overline{k}}^T (p \frac{A_{\overline{k}}}{\rho_{\overline{k}}} ) {\bf x}_{\overline{k}}
\end{equation}
For $\overline{k}$, since $A_{\overline{k}}$ is symmetry positive definite, $C_{\rho_{\overline{k}}} < 0$.

Next, we evaluate the absolute value of $C_{\rho_{\overline{k}}}$.
Using ${\bf x} = A^{-1}b$ and Eq.(\ref{yousogouseigyouretu}), $C_{\rho_{\overline{k}}}$ can be rewritten as follows:
\begin{eqnarray}
C_{\rho_{\overline{k}}} & = & - p \rho_{\overline{k}}^{-1} {\bf b}_{\overline{k}}^T A_{\overline{k}}^{-1} {\bf b}_{\overline{k}} \nonumber \\
& = & - p \rho_{\overline{k}}^{-(p+1)} {\bf b}_{\overline{k}}^T D_{\overline{k}}^{-1} {\bf b}_{\overline{k}}
\end{eqnarray}
Since $p$, ${\bf b}_{\overline{k}}$, and $D_{\overline{k}}$ are constant, we have
\begin{equation}
\lim_{\rho_{\overline{k}} \rightarrow 0} C_{\rho_{\overline{k}}} = - \infty
\end{equation}

Based on the above, we will show that the density $\rho_{\overline{k}}$ of the $\overline{k}$-th element adjacent to the $j$-th element ( $b_j \neq 0$ ) moves away from the value $0$ even if the initial guess of $\rho_{\overline{k}}$ is close to $0$.
$\rho_{\overline{k}}$ is obtained as a solution of constrained optimization problem by iterative methods in general.
Then we examine for the following three principal methods:
\begin{enumerate}
\item gradient vector based method including steepest descent method and CGM
\item the OC method
\item the CONLIN method
\end{enumerate}

First, for the gradient vector based method, 
since $\rho_{\overline{k}}$ is updated to the negative direction of $C_{\rho_{\overline{k}}}$,
$\rho_{\overline{k}}$ definitely increases and does not converge to zero.

Next, we examine for the OC method.
The updating equation of $\rho_{\overline{k}}$ is as follows\cite{Fujii:Suzuki:Ohtsubo2000}：
\begin{equation}
\rho_{\overline{k}}^{(t+1)} = \left( \frac{C_{\rho_{\overline{k}}}^{(t)}}{\lambda} \right)^{0.85} \rho_{\overline{k}}^{(t)}
\end{equation}
where $\lambda < 0$ is a Lagrange multiplier.
From the consideration of the case when $\rho_{\overline{k}}$ is close to zero,
there exists $\rho_{\overline{k}} > 0$ so that $\frac{C_{\rho_{\overline{k}}}^{(t)}}{\lambda} > 1$.
Since at that point $\rho_{\overline{k}}^{(t+1)} > \rho_{\overline{k}}^{(t)}$, the above updating equation moves $\rho_{\overline{k}}$ away from 0.

The updating equation in this case by the CONLIN method is derived from ref.\cite{Fleury1989} as follows:
\begin{equation}
\rho_{\overline{k}}^{(t+1)} = \left( - \frac{C_{\rho_{\overline{k}}}^{(t)}}{\lambda} \right)^{1/2} \rho_{\overline{k}}^{(t)}
\end{equation}
where $\lambda > 0$ is a Lagrange multiplier.
Similar to the OC method, 
From the consideration of the case when $\rho_{\overline{k}}$ is close to zero,
there exists $\rho_{\overline{k}} > 0$ so that $(-\frac{C_{\rho_{\overline{k}}}^{(t)}}{\lambda}) > 1$.
Since at that point $\rho_{\overline{k}}^{(t+1)} > \rho_{\overline{k}}^{(t)}$, the above updating equation moves $\rho_{\overline{k}}$ away from 0.

As described above, $\rho_{\overline{k}}$ does not converge to zero when solving by the iterative methods with the initial value of $0 < \rho_{\overline{k}} \le 1$.
Now the proposition-2 was proved and so was proposition-1.
But it must be noted that if $\rho_{k}$ is initialized to be $0$, the CGM breaks down at the time when calculating its sensitivity $C_{\rho_{k}}$.

\section{Simulation}
What is verified in computer simulations is that the CGM converges without breakdown to a local optimal solution even if the total stiffness matrix becomes singular.

One can see, however, that the convergence rate of the density to the value 0 is much slow while the rate to the value 1 is very fast both in the OC and the CONLIN method.
That is, the density going toward zero takes the tiny values such as $10^{-10}$, $10^{-20}$, and $10^{-30}$ endlessly.
Then the total stiffness matrix does not easily become singular.
Therefore, we add the operation to force the density value be $0$ if it is less than a predetermined value ( $10^{-3}$ in the following simulations ).
Besides, we use a simple pre-conditioning matrix for CG and CR method, every diagonal element of which are equal to the inverse of the corresponding diagonal element of the stiffness matrix.
Pre-conditioned CG and CR is denoted as PCG and PCR, respectively.

We compare the solutions obtained by the following four methods:
\begin{enumerate}
\item PCG-OC
\item PCR-OC
\item PCG-CONLIN
\item PCR-CONLIN
\end{enumerate}

The environment of the computer simulation is as follows:
\begin{itemize}
\item CPU: PentiumⅢ(933MHz)
\item memories : 512MB
\item compiler : gcc version 2.95.3 20010315 (release)
\item compiler option : O2
\end{itemize}
Some pre-determined values are as follows:
\begin{itemize}
\item upper bound of the total volume of the structure: $\rho_{MAX} / \sum_j \rho_j = 0.375$
\item convergence criteria of structural analysis problem: repeat until either the residual becomes less than $10^{-8}$, or the number of iterations exceeds a predetermined value.
The predetermined value is set to be the number of nodes.
\item convergence criteria of the OC or CONLIN method:
repeat until either the absolute value of the change of Lagrangian becomes less than $10^{-10}$, or  the number of iterations exceeds the predetermined value.
The predetermined value is set to be 100.
\end{itemize}

%
\subsection{Two bar truss problem}
\begin{figure}[hbtp]
  \begin{center}
  \includegraphics[scale=0.5]{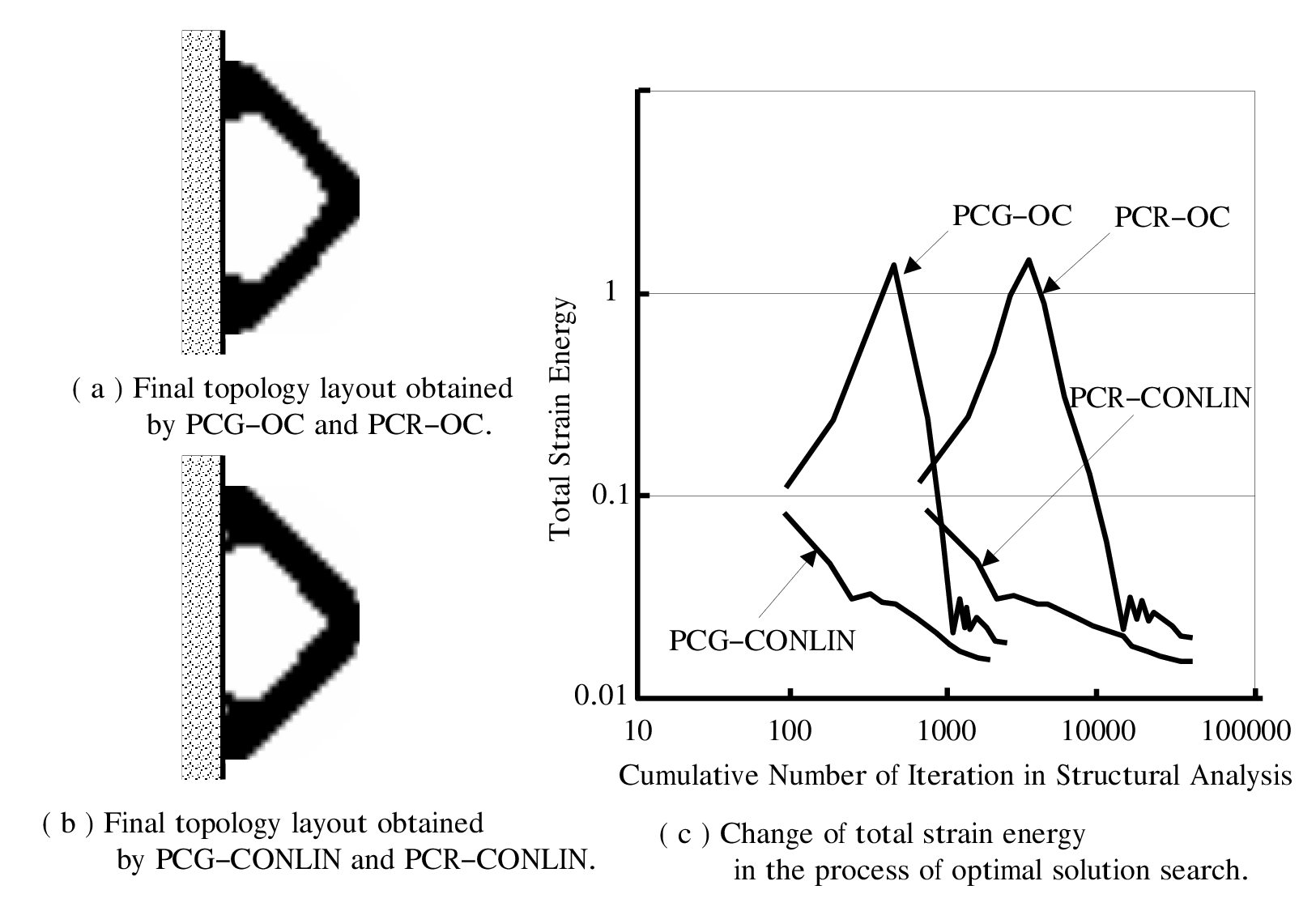}
    \caption{simulation results for two bar truss problem}
    \label{Fig2bartruss}
  \end{center}
\end{figure}

A simple problem for which the solution is obtained analytically is considered as a verification of the previously described proof.
Two bar truss problem is well known one of such problems.
Fig.\ref{Problem01}(a) illustrates the problem definition.
The design domain, $a=10[mm]$, $b=20[mm]$, is discretized using a $20 \times 40$ mesh of four-node bilinear plane strain elements.
The material parameters are assumed to be Young's modulus $E = 2.1 \times 10^5[N/mm^2]$ and Poisson's ratio $\nu = 0.3$.
The load at the middle of the free end is assumed to be $P = 1.05 \times 10^2[N]$.

The final topology layouts using the four methods are given in Fig.\ref{Fig2bartruss}(a) and (b).
All solutions give so-called two bar trusses with an internal angle of $90^{\circ}$ which is exactly the same as the analytical solution \cite{Rozvany:Bondsoe:Kirsh1995, Rozvany:Zhou:Sigmund1994}.

Fig.\ref{Fig2bartruss}(c) shows the total strain energy iterative histories for the four methods.
The horizontal and vertical axis in Fig.\ref{Fig2bartruss}(c) is the cumulative number of iterations of PCG or PCR methods and total strain energy, respectively, in log-log scale.
In the histories, we can see that PCG-CONLIN is the fastest among others.
Indeed, as denoted in Fig.\ref{table1}, PCG-CONLIN obtains the minimum total strain energy by the minimum CPU time among others.
In the previous part of this article, we showed the convergence of the CRM and the CGM for singular systems.
The computer simulation verified that both of them can use for the structural analysis in topology optimization even if the stiffness matrix becomes singular.
It also showed that the CGM is much effective than the CRM in terms of computational cost.

\begin{figure}
\begin{center}
\begin{tabular}{|c|r|r|r|} \hline
method     & total iterations & total strain energy & CPU time (sec) \\ \hline
PCG-OC     &           $2599$ &         $0.0191866$ &       $141.66$ \\ \hline
PCG-CONLIN &           $2014$ &         $0.0160593$ &       $111.78$ \\ \hline
PCR-OC     &          $45031$ &         $0.0191104$ &      $1958.74$ \\ \hline
PCR-CONLIN &          $40033$ &         $0.0160725$ &      $1590.87$ \\ \hline
\end{tabular}
\end{center}
\caption{Comparison between methods for two bar truss problem}
\label{table1}
\end{figure}

\section{Conclusion}
In topology optimization,
a singular stiffness matrix is often encountered because the densities of some elements of the structure become zero.
To avoid the numerical breakdown caused by this singularity,
the previous methods use some additional procedures for regularizing the singular matrix.
For example, the lower bound of the density was introduced for the material distribution method.
These procedures have, however, some problems.
To resolve such a singular system without additional procedures,
we focused on the convergence properties of the conjugate residual method (CRM) and the conjugate gradient method (CGM).
The convergence of the CRM for singular systems has been proved but has not been applied to the structural analysis in the previous work.
In this article, computer simulation showed that using the CRM as a solver of the structural analysis gives a local optimal solution in topology optimization.
Next, the convergence of the CGM for singular systems, especially in the case when the stiffness matrix becomes singular in topology optimization, considered in this article.
Although the idea that uses the CGM for solving the structural analysis in topology optimization has already been proposed, the proof of its convergence has not been given.
We proved the convergence of CGM when the stiffness matrix becomes singular.
It also holds for preconditioned CGM.
Computer simulations for an analytically solved problem verified our proof.
Because of no restriction for design variables,
the similar discussion can be available for the homogenization method.

\section{Acknowledgement}
The authors express their gratitude to Prof. Daiji Fujii of Kinki University, Prof. Ken Hayami of National Institute of Informatics and Prof. Shao-Liang Zhang of University of Tokyo for giving information related to their studies.

\end{document}